\documentclass{appolb}
\usepackage{graphicx}

\begin{document}
\title{Effect of confinement and stiffness on the conformational change of a semiflexible homopolymer chain...%
\thanks{pkmishrabhu@gmail.com}%
}
\author{Pramod Kumar Mishra
\address{Department of Physics, DSB Campus, Kumaun University, Nainital, Uttarakhand-263002 (INDIA)}
\\
}
\maketitle
\begin{abstract}
We analyse the nature of the confinement of an infinitely long (and finite) linear semiflexible homo-polymer chain confined in between two geometrical 
constraints ($A \& B$) under good solvent condition in two dimensions. The constraints are stair shaped impenetrable lines.
A lattice model of fully directed self avoiding walk is used to list information of walks of the confined  
chain and the exact enumeration technique is used for the canonical ensemble of conformations of the confined chain to 
discuss equilibrium statistics of the chain. We obtain the probability of polymerization of the confined flexible chain segments
with either one end (polymer trains) or both the ends of the confined chain lying on the stair shaped constraints
(polymer bridge and arc). We have also calculated the force of confinement exerted by the constraints on to the chain or the force 
exerted by the chain on the geometrical constraints using grand canonical ensemble theory and discuss nature of variation of the force.
\end{abstract}
  
\section{Introduction}
The soft macromolecule ($DNA$ \& $proteins$) is easily squeezed into much smaller space than the natural size of the molecule. 
For example, $actin$ filaments in $eukaryotic$ cell or $protein$ encapsulated in $Ecoli$ \cite{1,2} 
are well known confined molecules. The study of such confined biomolecules may serve as the basis 
for understanding molecular processes occurring in the living cells. 
Advances in the single molecule based experiments have made it possible to study conformational change of a confined macromolecule. 
Therefore, the study of the conformational properties and the possibility of polymerization of the polymer chain segments with either one end 
of the chain lying on the constraints (polymer train) or both the ends lying on the flat geometrical constraints 
(polymer bridge and polymer loop) have attracted attention in the recent years, see 
\cite{3,4,5,6,7,8,9,10,11,12,13}, and the references quoted therein. 
The polymer bridge \cite{13} corresponds to the confined chain segments with chain ends lying on two different 
geometrical constraints, however, arc corresponds to the chain segments with  both end monomers of the chain lying on a constraint.  
The polymer loop has end monomers in the vicinity and arc corresponds to chain segments with its end monomer not lying
in the vicinity.   

In this paper, we consider a long linear semiflexible homopolymer chain confined (as shown in figure-1) in between two stair shaped geometrical
constraints ($A$ and $B$). We model the chain as a directed self-avoiding walk ($DSAW$) \cite{8,9,10,14,15,16,17,18} and list
information of all the possible walks
of the confined chain of given length with the help of a square lattice. The information of the walks of the confined chain is used as
per the exact enumeration method \cite{19,20,21,22,23,24,25,26,27,28}. 
The ratio method of series analysis \cite{22} is used to discuss conformational properties
of the confined chain. Since, the directed self-avoiding walk model is solvable analytically \cite{14,15,16,17,18}, therefore,  
the exact values of $g_c$ \cite{8} of the fully directed self avoiding walk ($FDSAW$) model is used 
to calculate the force of confinement on the chain due to the constraints or on the constraints due to confined polymer chain.
The nature of variation of the force of confinement with the constraints separation is discussed.  
The separation between the geometrical constraints is increased in the unit of the monomer (or step) size of the chain (walk). 

\begin{figure}[htb]
\centerline{\includegraphics[width=0.5\textwidth]{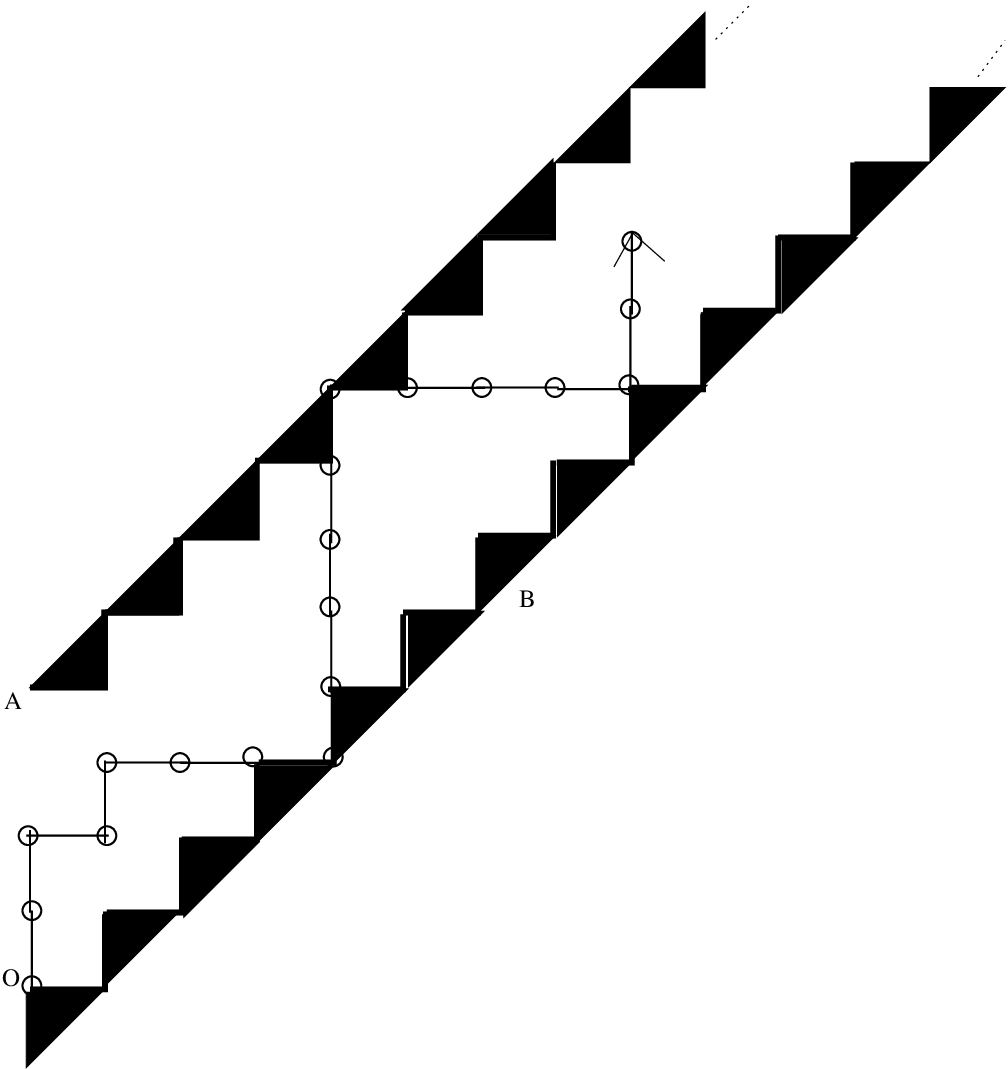}}
\caption{A walk of an infinitely long and linear semiflexible homo-polymer
chain is schematically shown in between two geometrical constraints ($A$ \& $B$). The 
separation ($n$) between the constraints is defined as the maximum 
number of steps that a walker can successively move either along $+x$ or
$+y$ direction. The value of $n$ is 5 for this figure.
} \label{Figure1}
\end{figure}
The paper is organized as follows: In Section 2, we consider a long linear semiflexible homopolymer chain confined in between
two geometrical constraints ($A$ and $B$) and fully directed self avoiding walk model is described for so 
confined polymer chain (as shown in figure-1) to collect information of confined chain walks (conformations). 
We have also described in brief the exact enumeration method for the proposed model of the confined chain, in addition 
to description of the method used to calculate the force of confinement and the exponent. In Section 3, 
data of the confined finite chain walks is used to discuss the possibility of the polymerization of the confined flexible polymer 
chain segments. The segments of the confined chain
are polymer train, bridge and arc. The role of the confinement and the stiffness on the conformational change of the chain in discussed in the section-4.
We have calculated the force of confinement that was exerted by the geometrical constraints on the polymer chain or the 
force exerted by the chain on geometrical constraints  and discuss nature of variation of the force in the section-5.
Finally, in Section 6, we summarize and discuss the results obtained.
\section{Model and method}
A lattice model of fully directed self-avoiding walk \cite{14,15} is used to study the coformational behaviour of a semiflexible
homopolymer chain confined in between two geometrical constraints ($A \& B$). The walks of the confined chain start from a point, $O$, lying
on the constraint $B$ (as shown in figure-1). However, in the case of collection of the data for the chain segments (polymer train, bridge and arc) 
the confined chain is grafted on the constraint $B$ at another point such that the first step
that the walker can take is along +y direction only ({\it i. e.} only one monomer of an end of the chain may lie on the constraint). 
In other words, the point $O$ is shifted to one monomer (step) along $+x$ direction for the
collection of the data for chain segments (train, bridge and arc). 
In this situation, we are able to obtain confined chain segments having either one end of the chain on the constraint or both the
ends of the chain on the constraint (or constraints). 
When only one end of the chain is one the constraint, the segment (conformation) of the chain is named as
polymer train \cite{13}. The polymer bridge \cite{13} corresponds to the chain conformations with one end lying on the constraint (say, $B$) 
while the other end on the another constraint (say, $A$).
However, the chain conformations with both the ends lying on the same constraint and at different points in the
confined space correspond to polymer arc.   

In the present investigation we consider directed self avoiding walk and the walker is allowed to take steps along $+x$ and $+y$ 
directions of the space available in between the two impenetrable constraints, {\it i. e.} the walker's steps are fully directed.
The fully directed self-avoiding walk model is restrictive in a sense that the angle of bend is either $90^{\circ}$ or there is no
bend. However, such model can be solved analytically and therefore gives exact results \cite{8,9,10,14,15,16,17,18,26,27,28}. 
It is possible to enumerate walks of relatively larger chain size in a directed walk model than the corresponding isotropic self-avoiding walk model.
Therefore, we have chosen fully directed self avoiding walk to model the chain in between the constraints and to list the information
of confined chain conformations as per the exact enumeration method. 

The stiffness of the confined chain is accounted by associating a Boltzmann weight with bending energy for each turn in the walk of the polymer chain. 
The stiffness weight is $k${\Large[}$=exp(-\beta\epsilon_{b})$; where $\beta=\frac{1}{k_bT}$ is 
inverse of the temperature, $\epsilon_b(>0)$ is the energy associated with 
each bend in the walk of the chain, $k_b$ is the Boltzmann constant and $T$ is temperature{\Large]}. 
For $k=1$ or $\epsilon_{b}=0$ the chain is said to be flexible and for 
$0<k<1$ or $0<\epsilon_{b} <\infty$ the polymer chain is said to be 
semiflexible. However, when $\epsilon_{b}\to\infty$ or $k\to0$, 
the confined polymer chain has shape like a rigid rod \cite{8}.

The walks of the $N$ monomers long confined chain is enumerated with the help a square lattice. The information of all the possible conformations 
of $N$ monomers (where, $N=1,2,3,...,100$) long confined chain is listed as per the exact enumeration method \cite{19,
20,21,22,23,24,25,26,27,28}.
The separation ($n$) between the constraints is varied ($n$=3,4,5,...,10) in the unit of monomer (step) size, 
and the list of $C_N(n,N_b)$ {\it i. e.} the number of all the possible conformations
of the chain for all possible values of $N_b$ is listed for the polymer chain length, $N$.
The canonical partition function [$Z^C_N(n,k)$] of the confined semiflexible polymer chain of $N$ monomers is calculated using equation,
\begin{equation}
Z^{C}_N(n,k)=\sum_{N_b=0}^{N-1}C_N(n,N_b) k^{N_b},
\end{equation}
where, $C_N(n,N_b)$ is the number of walks (conformations) of the confined chain with 
number of bends ($N_b$) in the chain, the total number of monomers in
the chain is $N$ and the separation between the constraints ($A \& B $) is $n$. 
We obtain the value of thermodynamic quantities of the confined flexible chain by substituting $k$ equal to unity.
The free energy of an infinitely long chain confined in between the constraints (as shown in figure-1) is calculated using the ratio method
of series analysis \cite{22}.
The application of the ratio method for present study is based on an assumption that the total number of conformations [$Z^C_N(n,k)$] 
of the confined chain increases with the number of monomers ($N$) in the chain as $\sim {\mu^N(n,k)}{N}^{\gamma-1}$.

However, the grand canonical partition function [$Z^G_n(g,k)$] of the confined chain (as shown in figure-1) is obtained using equation,

\begin{equation}
Z_n^{G}(g,k)={\sum}^{N=\infty}_{N=0}Z^{C}_N(n,k) {g^N},
\end{equation}
where, $g$ is the step fugacity of each monomer of the confined chain \cite{8}. The critical value of step fugacity of the chain is determined from
singularity of the partition function $Z^G_n(g,k)$.

In this paper, we use the value of $g_c(n,k)$ from ref. \cite{8} 
$(3\le n\le 19)$ to calculate the
free energy of the confined semiflexible chain. The free energy so obtained is used to find the confining force exerted by the constraints on
each monomer of the chain or the force exerted on to the constraints due to each monomer of the confined semiflexible chain. 
The free energy per monomer of the confined semiflexible chain is calculated using well known formula,
\begin{equation}
A(n,k)=k_bTln[g_c(n,k)],
\end{equation}
The force of the confinement per monomer $f(n,k)$  
is evaluated for separation, $n$, between the constraints using relation,
\begin{equation}
f(n,k)=-\frac{\partial A(n,k)}{\partial n}=-  \frac{A(n+1,k)-A(n-1,k)}{(n+1)-(n-1)},
\end{equation}
We have calculated the universal exponent $\gamma$ by plotting a straight line,

\begin{equation}
ln[Z^C_{N+2}(n,k)]-ln[Z^C_N(n,k)]=2ln[\mu(n,k)]-(\gamma-1)[ln(N+2)-ln(N)],
\end{equation}
for different values of $n$ and the slope of the line is zero for the confined flexible polymer chain.
\section{The confined flexible polymer chain segments: Train, Bridge and Arc}
We collect the information of the confined polymer chain conformations of length $N$(=1,2,..100) and  
having either one end or both the ends of the chain conformations lying on the constraint (or constraints). This information is taken 
into account to find the exact number ($T_N^n$) of the flexible polymer train, the exact
number of bridge ($B_N^n$), the exact number of polymer arc ($L_N^n$), (where $n$ is the separation in between the constraints) 
and the number of all the conformations $C_N^n$ (Canonical partition function) 
of the confined flexible polymer chain. The ratio of the number of polymer segments (arc, bridge and train) to the canonical partition
function is nothing but the probability of the polymerization of the polymer segments ({i. e. } train, bridge and arc). 
The probability of the polymerization of polymer train of $N$ monomers
is shown in the figure-2, while the probability of the polymerization of the polymer bridge and the arc are shown in the figure-3.
\begin{figure}[htb]
\centerline{\includegraphics[width=0.80\textwidth]{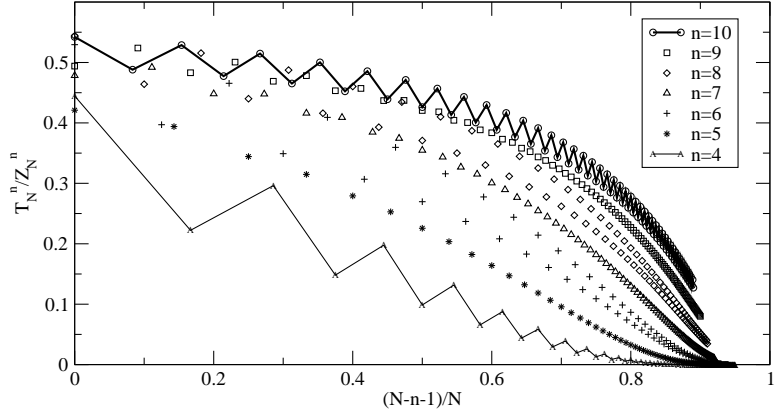}}
\caption{ The probability of the polymerization of the flexible polymer train is shown in this figure. 
The length of the chain ($N$) is varied from $N=1,2..$ to 100,
and the separation ($n$) between the fluctuating constraints is varied from 4 to 10 unit. There is odd-even fluctuation in the probabilty of polymerization
of polymer train for the given length of the confined polymer chain. 
} \label{Figure2}
\end{figure}

\begin{figure}[htb]
\centerline{\includegraphics[width=.80\textwidth]{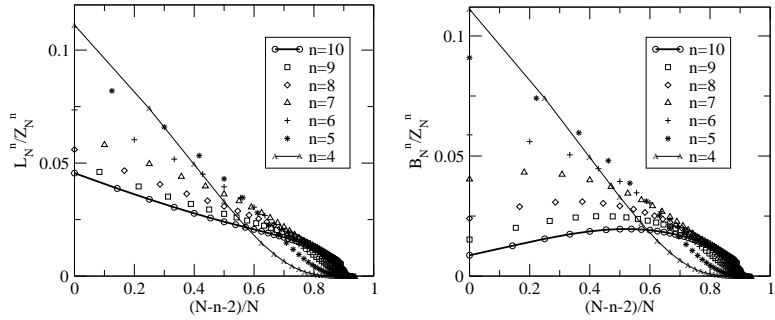}}
\caption{We have shown the probability of polymerization of polymer bridge and polymer arc 
for the flexible chain confined by fluctuating impenetrable constraints ($A$ and $B$). 
} \label{Figure3}
\end{figure}
\section{The statistics of the confined semiflexible polymer chain}
The critical value of effective coordination number $\mu_c(n,k)=\frac{1}{g_c(n,k)}$ 
is calculated using zeroth order approximation of the ratio method 
\cite{19,20,21,22,23,24,25,26,27,28}.
Our result for the critical value of step fugacity in the zeroth order approximation is 0.61803 for the confined 
flexible chain and the value of $n$ is equal to 3, and this value of step fugacity is 
in the excellent agreement with our exact result 0.61803 \cite{8}. However, zeroth order estimate of $g_c(n,k)$ is not in good agreement with 
our exact results for the stiffer chains \cite{27}. It appears that the number of conformations [$Z_N^C(n,k)$] 
are small and it is not increasing according
to relation $\sim {\mu^N(n,k)}{N}^{\gamma-1}$ for the confined stiffer chain.
We have shown the variation of the dominant contribution of the canonical partition function of the confined semiflexible polymer chain for different  
number of bends per monomer ($N_b/N$) in the figure-4. It has been found that the canonical
partition function of the confined flexible polymer
chain is dominated by walks having number of bends per monomer 0.606 (as shown in figure-5) 
and the separation between the constraints is four
unit. The dominant contribution of the partition function shifts to a smaller value of the number of bends per monomer of 
the confined flexible chain as the spacing between the constraints is increased. For example, when $n$ is equal to 10 
the peak of the curve (as shown in figure-4)
shifts to ($N_b/N$=) 0.526 for the confined flexible chain. 
It has also been found that the peak of the curve indicating dominant contribution of the partition function (as shown in figure-4) 
shifts to a smaller value of number of bends per monomer of the confined chain
for the given separation between the constraints, $n=4$, as the stiffness ($k$) of the confined chain varies from 1 to 0.4. For example, the
peak position shifts from 0.606 to 0.456 for the confined chain as $k$ is altered from 1 to 0.4 for the separation between the constraints 
equal to four unit.     
\begin{figure}[htb]
\centerline{\includegraphics[width=1.0\textwidth]{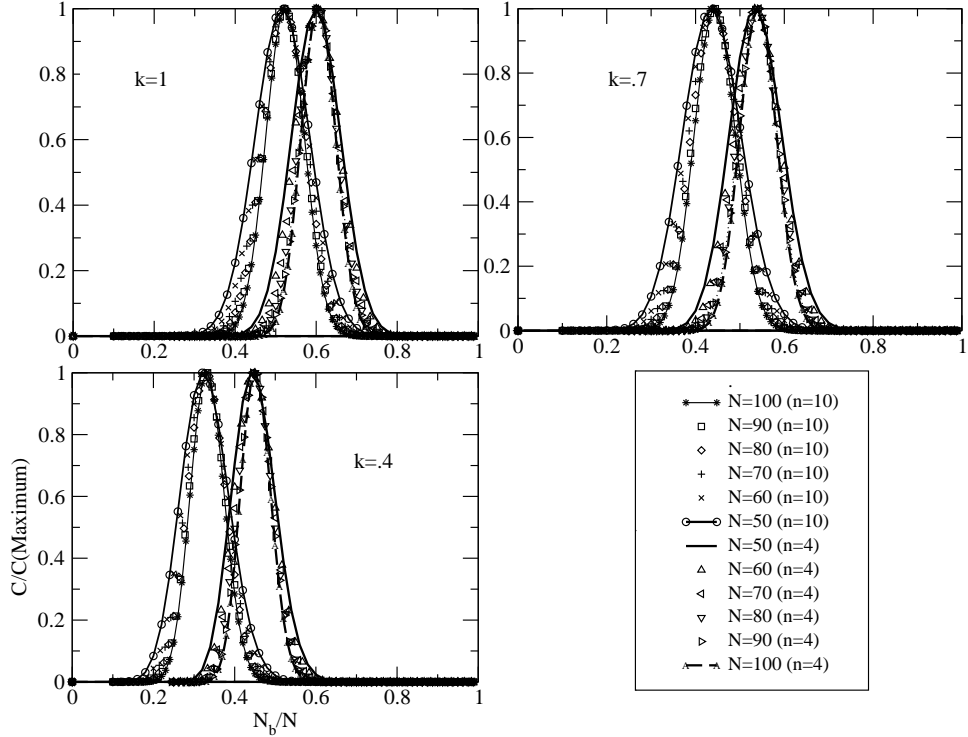}}
\caption{ The dominant contributions of the partition function are shown in this figure 
for different values of the number of bends per monomer in the confined polymer chain. It is  
shown that the peak position shifts with the constraints separation $n$  (from 4 to 10) or with the stiffness ($k$) of the confined 
polymer chain (from 1 to 0.4) monotonously.  
} \label{Figure4}
\end{figure}

\section{The force of the confinement}
The free energy of the chain is written in the unit of thermal energy $k_bT$(=25.875meV) per monomer 
at room temperature ($T$=300K). We use the value of step length (monomer size) 
4.14 \AA. So that the force can be expressed in the unit of 10$pN$. Since, the force of confinement is of the order  
$pN$ and, therefore, the force may be measured with the help of $AFM$.
The nature of variation of the confining force per monomer [$f(n,k)$] with the constraints separation ($n$) is a 
curve so that $f(n,k)*n^{q(k)}$=constant, where $q(k)$ is a fraction and the value of $q(k)$ depends on the stiffness of the chain. 
Therefore, the log-log plot of $f(n,k)-n$ is a straight line
for the confined flexible ($k=1$) and the confined semiflexible chain ($k$= (0.7 and 0.4), as shown in the figure 6). 
The slope of $ln[f(n,k)]$-$ln[n]$ curve is 
varying gradually from -2.5189 (for the confined flexible chain) to -2.4266 for the confined semiflexible chain ($k$=0.4).

Finally, in figure 7, we have shown the variation of $\mu_c=\frac{1}{g_c(n,k)}$ with $n$ to discuss the role of the stiffness $k$ and the confinement 
on the conformational statistics of the chain. 
\begin{figure}[htb]
\centerline{\includegraphics[width=0.8\textwidth]{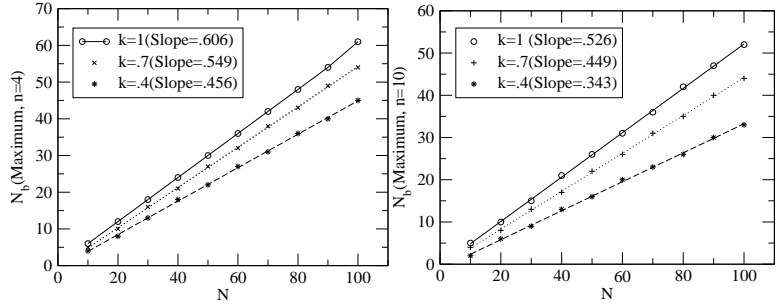}}
\caption{
We have shown the variation of the number of bends of the confined chain corresponding to a dominant contribution of the partition function. The
slope of the curve varies with the stiffness of the confined chain. The slope of the curve also varies with the variation of the 
constraints separation ($n$).}
\label{Figure5}
\end{figure}

\section{Summary and conclusions}
We consider a long linear semiflexible homopolymer chain confined in between two ($A$\&$B$) impenetrable stair shaped 
geometrical constraint (surface) under good solvent condition (as shown in figure-1). We use fully directed self avoiding walk model on a square lattice
to model the situation of the confinement. The information of all the possible walks of the finite size (N=1,2,3, ..,100) chain is
listed as per the exact enumeration method \cite{19,20,21,22,23,24,25,26,27,28} and the information of the walks (chain conformations) is 
used to calculate the thermodynamic properties and the conformational change of the confined polymer chain.
In the case of calculation of probability of polymerization of the confined chain segments (polymer arc, bridge \cite{13} and 
polymer train \cite{13}) the chain is grafted on the constraint $B$ at another point shifted by one step along $+x$ direction from $O$
so that only one monomer (at the end where chain is grafted on the constraint) may lie on the constraint.

\vspace{.55cm}
\begin{figure}[htb]
\centerline{\includegraphics[width=0.7\textwidth]{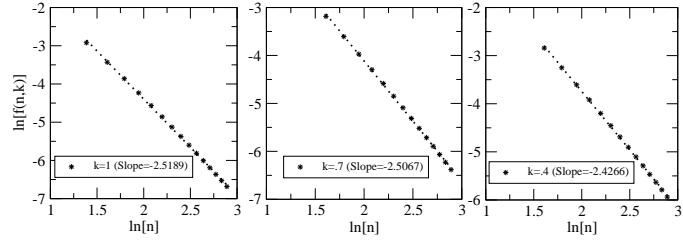}}
\caption{We have plotted the logrithm of the confining force [$f(n,k)$] with 
logrithm of the separation ($n$) in-between the constraints for the confined flexible ($k=1$) and the confined 
semiflexible chain ($k$=0.7, and 0.4). The log-log curve is linear for the chosen values of $k$. The log-log plot of
[$f(n,k)-n$] is shown for a flexible chain ($k=1$) and semiflexible chain ($k$=0.7 and 0.4). The force is reported
in the unit 10$pN$ at room temperature 300$K$ for monomer (step) size 4.14\AA.}
\label{Figure6}
\end{figure}

The probability of polymerization of the flexible chain segments with one end lying on the fluctuating (stair shaped)
impenetrable constraint $B$ ({\it i. e.} the polymer train) is shown in the figure-2 and 
the probability of polymerization of the polymer bridge and the arc is shown in the figure-3. 
The probability of polymerization of the polymer train has odd-even fluctuation due to presence of geometrical constraint 
$B$. The odd-even fluctuation is well known in the literature of polymer statistics and to avoid effect of this fluctuation
we calculate the coordination number using relation $\mu=\sqrt(Z_{N+2}/Z_N)$ rather than $\mu=Z_{N+1}/Z_N$ in the ratio method.
The probabilty of polymerization of the polymer train 
reduces with effective length of the train. The probability of polymerization of the polymer arc also reduces with effective length 
of the chain arc.However, the probability of polymerization of 
the polymer bridge has one peak. In other words the peak position for polymer bridge is observed for an intermediate value of effective length
of the chain bridge. It is understood that the polymer train is the most probable polymer segments while the polymer bridge is the least probable
polymer segments.

It has been found that the conformation of the confined polymer chain gets modified as the separation between the constraints is varied
or the stiffness of the chain is altered (as shown in figure-4). The canonical partition function of the confined flexible 
chain is dominated by walks having number of bends $N_b=0.606*N$, provided, the separation between the constraints is four unit (as shown in
figure-5). The dominant contribution of the canonical partition function corresponds to the chain conformations having value of 
$N_b=0.526*N$, as the constraints separation is increased to ($n=$) 10. However, the canonical
partition function of the confined chain is dominated by walks (conformations) having $N_b=0.456*N$ as the 
stiffness of the chain is modified to $k=0.4$ and the value of constraints separation is ($n=)4$ (as shown in figure-5). 
The confined semiflexible and flexible chain conformations are modified with simple scaling (linear) due to sever restrictions
imposed on the chain.

The step fugacity $(g)$ is the reciprocal of the lattice coordination number $(\mu)$ of the step that a walker can take while
enumerating conformation of the polymer chain.
The ratio method \cite{22} is used to obtain the critical value of step fugacity [$g_c(n,k)$] of the confined chain in
the thermodynamic limit. The critical value of step fugacity ($g_c$) for the confined semiflexible chain is found in very good
agreement with our analytical results for the confined chain \cite{8}. 
However, the ratio method fails for short and stiffer chains \cite{27} and therefore, the exact values of $g_c(n,k)$ \cite{8} is used to 
determine the free energy per monomer [$A(n,k)$] of the chain. 
The value of the free energy is used to calculate the force of the confinement [$f(n,k)$] exerted by each monomer on the constraints or
the force of confinement experienced by each monomer due to presence of constraints around the chain. The value
of spacing ($n$) $(4\le n\le19)$ is altered in the unit of monomer (step) size.

Since, variation of force of confinement ($f(n,k)$) with $n$ is such that the curve $f(n,k)$-$n$ follows a condition, $f(n,k)*n^{q(k)}$=constant, 
[where, $q(k)$ is nothing but the slope of curve that is shown in the figure (6)], and, therefore the log-log
plot of $f(n,k)$-n is a straight line, unlike our previous findings \cite{24}. The deviation from earlier reported result 
is due to severe restrictions imposed
on the chain. The restrictions are geometrical constraints, the stiffness of the chain and the directedness of the chain.
The slope of the straight line (as shown in figure-6) varies gradually with the variation of stiffness of the chain. 
The curve is a straight line for a flexible chain and 
with increase of stiffness of the chain, the slope of the curve decreases. The slope of the curve $f(n,k)$-$n$ is -2.5189
for confined flexible chain and vlaue of the slope differs from -1.66 \cite{29} for a confined worm like chain of length $L$
due to different geometrical restrictions imposed on an infinitely long chain. 
 There is large entropy [$\sim \mu_c(n,k)^N$] of 
the confined flexible chain and due to the confinement,
the entropy is reduced. The entropy of the chain is increased in the controlled manner with the increase of separation ($n$)
and therefore, log-log plot of $f(n,k)$-$n$ is a straight line. 

When there is no constraints around the chain, the entropy of the flexible chain increases with the chain size ($N$) as $\sim 2^N$. 
However, for the stiffer chain
the entropy increases as $\sim (1+k)^N$ \cite{26}. The value of $k$ is less than 1 for the stiffer chains. 
The constraints around the chain further reduces 
the entropy and the entropy increases with the chain size as $\sim (\mu_c)^N$ (where 
$\mu_c<(1+k)$, {\it i. e.} $\mu_c$(bulk) as shown in figure-7).   

\begin{figure}[htb]
\centerline{\includegraphics[width=0.8\textwidth]{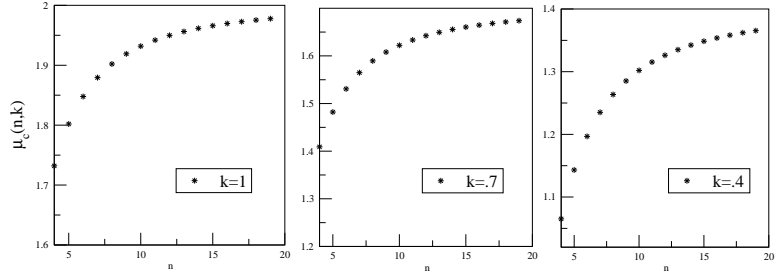}}
\caption{ The critical value of effective co-ordination number of the confined chain is shown for the flexible ($k$=1) and the
semiflexible ($k$=0.7 and 0.4) confined chain in this figure and the separation ($n$) between the constraints ($A$ \& $B$) is varying from 4 to 19.   
However, in the bulk (when there is no geometrical constraints near the chain) 
the effective co-ordination number is =1+$k$, \cite{26}.
} \label{Figure7}
\end{figure}

It appears form our present investigation that the log-log plot of $f(n,k)$-$n$ may deviate from straight 
line for possible stiffer chain that may polymerize 
in between the constraints. This is due to the fact that the entropy of the stiffer chain is much smaller than the flexible chain and presence 
of the constraints around the chain further reduces entropy of the stiffer chain. Therefore, the entropy is radially reduced to a small value
and the curve may deviate from the straight line ({\it i. e.} for $\mu\to$ 1).

We would like to comment about the value of the exponent $\gamma$ for the confined homopolymer chain. It has been found for the confined 
flexible chain (as shown in figure-1) that the value of the exponent $\gamma$ is unity. However, the method used in this paper for
determination of the exponent is not suitable for smaller number of conformations of the short chains.  

\section{acknowledgements}
 The financial support received from SERB, Department of Science and Technology, New Delhi (SR/FTP/PS-122/2010) thankfully acknowledged.



\end{document}